\documentclass[aps,pra,twocolumn,showpacs]{revtex4}

\usepackage{amsmath}
\usepackage{amsfonts}
\usepackage{amssymb}
\usepackage{graphics}
\usepackage{graphicx}

\begin{document}


\title{Extrinsic orbital angular momentum of entangled photon-pairs in \\spontaneous parametric down-conversion}

\author{Sheng Feng} \email{sfeng@ece.northwestern.edu}
\author {Prem Kumar}
\affiliation{Center for Photonic Communication and Computing, EECS Department, Northwestern University, Evanston, IL 60208-3118, U.S.A.}




\date{\today}

\begin{abstract}
Starting from the standard Hamiltonian describing the optical non-linear process of spontaneous parametric down-conversion, we theoretically show that the generated entangled photon-pairs carry non-negligible orbital angular momentum in the degrees of freedom of relative movement in the type-II cases due to spatial symmetry breaking. We also show that the orbital angular momentum carried by photon-pairs in these degrees of freedom escapes detection in the traditional measurement scheme, which demands development of new techniques for further experimental investigations.
\end{abstract}


\maketitle

\section{\label{sec:intro}Introduction}

The nonlinear optical process of spontaneous parametric down-conversion (SPDC) serves as an important toolkit to produce entangled photon-pairs. The entanglement may involve energy, linear momentum, and angular momentum. Recently, it was demonstrated that photon-pairs generated in SPDC processes are entangled in another physical variable, orbital angular momentum (OAM), in which the states of high-dimensional entanglement \cite{mair01} and hyper-entanglement \cite{barreiro05} can be prepared.
 
Although OAM of light is not a true angular momentum \cite{enk94,arnaut00}, it {\em is} a measurable physical quantity under paraxial approximation. Therefore, the total angular momentum of light may be broken into three separate measurable parts: spin angular momentum, intrinsic OAM, and extrinsic OAM \cite{bliokh06}. The spin of light is determined by the polarization of the light beam, the intrinsic OAM of light is associated with the transverse phase fronts of light beams \cite{allen92}, and extrinsic OAM of light is related to the relative movement of the center of the light beam with respect to some external point in space. As one shall see below, all previous investigations on OAM entanglement generated in SPDC processes are limited to intrinsic OAM of light. However, we show in this paper that the down-converted photon-pairs in type-II SPDC processes carry non-negligible extrinsic OAM.

The extrinsic OAM carried by photon-pairs in SPDC processes is the key to understand the relation of spatial symmetry to the OAM conservation rule in SPDC processes \cite{feng08}. Moreover, our study may lead to a new path to exploit the orbital angular momentum of light in practical applications. In section II, we introduce the states of light carrying OAM and derive the mathematical descriptions for intrinsic and extrinsic OAM of light. Section III calculates the traditional entangled states of light in OAM created in SPDC processes, which are states entangled in intrinsic OAM. We further show that, as a result of azimuthal symmetry breaking, photon-pairs created in type-II SPDC processes carry non-negligible extrinsic OAM, which is beyond detection scope of traditional scheme \cite{mair01,Oemrawsingh05}. Finally, brief discussions are given in section IV followed by a summarizing conclusion.

\section{\label{sec:oam}OAM of light}

When the conservation of a vector is concerned, one usually breaks the vector into components along, for example, $x$, $y$, and $z$-axes. The conservation of the vector means all these components are conserved. If one of the components is not conserved, the vector is said to be non-conserved. The same argument applies to the cases of OAM (non-)conservation in the SPDC processes. To exam whether the OAM is conserved along the pump propagation direction ($z$-direction) in the SPDC processes, one needs to calculate the state of the down-converted light beams that carry OAMs, the $z$-components of which are, say, $L_z^{(s)}$ and $L_z^{(i)}$. 

An elegant approach exploiting orbital Poincar\'{e} spheres to study this general case was provided in \cite{calvo06}, where the $z$-component of the OAM carried by photons is, however, not guaranteed to be $n\hbar$ ( $n$ is an integer). We quantize the studied field along $z$-axis such that the $z$-component of the OAM of light is $n\hbar$ ( $n$ is any integer). In quantum theory, the eigenstate of the OAM operator $\hat{L}_z$ can be found by solving the eigenvalue equation:
\begin{equation}
\hat{L}_z|\psi_1(t)\rangle=\alpha|\psi_1(t)\rangle, \nonumber
\end{equation}
which leads to a solution for a one-photon field in free space as follows \cite{arnaut00},
\begin{equation}
|\psi_1(t)\rangle=
\sum\limits_{\bf k} \ g({\bf k},t)e^{il\phi_{\bf k}}\hat{a}^\dagger({\bf k})|0\rangle.
\label{psi1}
\end{equation}
Here $\alpha=l \hbar$ is the $z$-component of the OAM carried by the photon ($l$ is any integer), and $\phi_{\bf k}$ the azimuthal angle of the wave vector ${\bf k}$. The $g({\bf k},t)$ is a function {\em independent} of the azimuthal angle $\phi_{\bf k}$ and, therefore, can be written in the form $g(p_\rho, k_z, t)$, where $p_\rho=\sqrt{{\bf k}^2-k_z^2}$ is the amplitude of the transverse component ${\bf p}_\rho$ of ${\bf k}$. The $\hat{a}^\dagger({\bf k})$ is the photon creation operator. We note that the freedom of the field polarization is neglected and emphasizes that the $l \hbar$ in Eq. (\ref{psi1}) has a physical meaning (OAM along the $z$-axis) essentially different from what is meant by the same notation in \cite{arnold02,walborn04,terriza03}, where the $l \hbar$ represents the OAM carried by photons along the axes dictated by the light-beam central vectors.

The one-photon detection amplitude $\varphi_1^l({\bf r})\equiv\langle0|\hat{E}^{(+)}({\bf r})|\psi_1(t)\rangle$ [$\hat{E}^{(+)}({\bf r})=\sum_{\bf k}C_k\hat{a}({\bf k})e^{i({\bf k}\cdot{\bf r}-\omega_kt)}$, $\hat{a}$({\bf k}) is the annihilation operator, $C_{k}$ is a coefficient dependent of $k=|{\bf k}|$] for the one-photon field in the eigenstate of the operator $\hat{L}_z$ is
\begin{eqnarray}
\varphi_1^l({\bf r})=
\sum\limits_{\bf k} \ g(p_\rho,k_z,t)C_{k}e^{il\phi_{\bf k}}e^{i({\bf k}\cdot{\bf r}-\omega_k t)}. 
\nonumber
\end{eqnarray}
With beam invariants ${\bf p}_\rho={\bf k}-k_z\hat{z}$ and $\omega$ introduced \cite{rubin96}, in a plane ($z=z_0$) transverse to the $z$-axis, the one-photon detection amplitude is then
\begin{equation}
\varphi_1^l({\bf q}_\rho)=
\sum\limits_{{\bf p}_\rho} \ h(p_\rho,t)e^{il\phi_{p}}e^{i{\bf p}_\rho\cdot{\bf q}_\rho},
\end{equation}
where $h(p_\rho,t)=\sum_\omega C_k g(p_\rho,k_z,t)e^{i(k_zz_0-\omega t)}$, in which $k_z=\sqrt{(\omega/c)^2-p^2_\rho}$, ${\bf q}_\rho={\bf r}-z\hat{z}$, and $\phi_{p}=\phi_{\bf k}$. 

We need to point out that the total OAM, the $z$-component of which is $L_z=l\hbar$ ($l$  is any non-zero integer), carried by a photon will be arbitrarily fractional if  the photon propagates along an arbitrary direction. Extensive theoretical study on fractional OAMs can be found in \cite{kastrup06}.

Similarly, the eigenstate of the OAM operator $\hat{L}_z^{(s)}+\hat{L}_z^{(i)}$ can be found by solving the eigenvalue equation
\begin{equation}
(\hat{L}_z^{(s)}+\hat{L}_z^{(i)})|\psi_2(t)\rangle=\alpha|\psi_2(t)\rangle, \nonumber
\end{equation}
which leads to a solution for two one-photon fields in free space as follows,
\begin{eqnarray}
|\psi_2(t)\rangle&=&
\sum\limits_{{\bf k}_s,{\bf k}_i} g_s(p_s,k_{z,s},t)g_i(p_i,k_{z,i},t)\nonumber \\
&\times&e^{i(l_s\phi_{{\bf k}_s}+l_i\phi_{{\bf k}_i})}\hat{a}_s^\dagger({\bf k}_s)\hat{a}_i^\dagger({\bf k}_i)|0\rangle.
\label{psi2}
\end{eqnarray}
Here $\alpha=(l_s+l_i) \hbar$ is the $z$-component of the OAM carried by the photons ($l_{s,i}$ are any integers) in both fields, and $\phi_{{\bf k}_{s,i}}$ the azimuthal angles of the wave vectors ${\bf k}_{s,i}$. In planes ($z_s=z_{s,0}, z_i=z_{i,0}$) transverse to the $z$-axis, the two-photon detection amplitude should read
\begin{eqnarray} \label{tpda2p}
\varphi_2^l({\bf q}_s,{\bf q}_i)&=&
\sum\limits_{{\bf p}_s,{\bf p}_i} h_s(p_s,t)h_i(p_i,t)e^{i(l_s\phi_s+l_i\phi_i)}\nonumber \\
&\times&e^{i({\bf p}_s\cdot{\bf q}_s+{\bf p}_i\cdot{\bf q}_i)},
\end{eqnarray}
where $\phi_{s,i}=\phi_{{\bf k}_{s,i}}$, and $h_{s,i}(p_{s,i},t)$ are functions independent of the azimuthal angles $\phi_{s,i}$. As we shall show below, the traditional scheme \cite{mair01,Oemrawsingh04ao} measures the OAM of the down-converted beams in SPDC processes in the degrees of freedom of center-of-momentum movement described by ${\bf q}_+={\bf q}_s+{\bf q}_i$ and ${\bf p}_+={\bf p}_s+{\bf p}_i$. Thereby, it is useful to re-write Eq. (\ref{tpda2p}) in terms of joint variables ${\bf q}_+$, ${\bf p}_+$, ${\bf q}_-={\bf q}_s-{\bf q}_i$, and ${\bf p}_-={\bf p}_s-{\bf p}_i$:
\newcommand{\bin}[2]{
\left( 
      \begin{array}
             {@{}c@{}}
             #1 \\ #2
      \end{array} 
\right)
}
\begin{eqnarray} \label{tpda2pp}
&&\varphi_2^l({\bf q}_+,{\bf q}_-)[\equiv \varphi_2^l({\bf q}_s,{\bf q}_i)]\nonumber\\
&=& \sum\limits_{{\bf p}_+,{\bf p}_-} \ h_2({p}_+,{p}_-,\phi_+-\phi_-,t)\nonumber\\
&\times&2^{-(l_s+l_i)}\left(p_+e^{i\phi_+}+p_-e^{i\phi_-}\right)^{l_s}\left(p_+e^{i\phi_+}-p_-e^{i\phi_-}\right)^{l_i} \nonumber\\
&\times& e^{i/2({\bf p}_+\cdot{\bf q}_++{\bf p}_-\cdot{\bf q}_-)}\nonumber\\
&=&\sum\limits_{{\bf p}_+,{\bf p}_-} \sum\limits_{m}h_2^{(m)}(p_+,p_-,t)e^{im(\phi_+-\phi_-)}\nonumber\\
&\times&2^{-(l_s+l_i)}\sum\limits_{n_s}\bin{l_s}{n_s}p_+^{n_s}e^{in_s\phi_+}p_-^{l_s-n_s}e^{i(l_s-n_s)\phi_-}\nonumber \\
&\times&\sum\limits_{n_i}\bin{l_i}{n_i}p_+^{n_i}e^{in_i\phi_+}(-p_-)^{l_i-n_i}e^{i(l_i-n_i)\phi_-}\nonumber \\
&\times&e^{i/2({\bf p}_+\cdot{\bf q}_++{\bf p}_-\cdot{\bf q}_-)}\nonumber \\
&=&\sum\limits_{m,n_s,n_i} 2^{-(l_s+l_i)}(-1)^{l_i-n_i}\bin{l_s}{n_s}\bin{l_i}{n_i}\nonumber\\
&\times&\sum\limits_{{\bf p}_+,{\bf p}_-} h_2^{(m)}(p_+,p_-,t)p_+^{n_s+n_i}p_-^{(l_s+l_i)-(n_s+n_i)}\nonumber \\
&\times&e^{i(m+n_s+n_i)\phi_+}e^{i[(l_s+l_i)-(m+n_s+n_i)]\phi_-}\nonumber \\
&\times&e^{i/2({\bf p}_+\cdot{\bf q}_++{\bf p}_-\cdot{\bf q}_-)},
\end{eqnarray}
where $\phi_\pm$ are the azimuthal angles of vectors ${\bf p}_\pm$ and exploited were $p_{\pm}e^{i\phi_\pm}=p_se^{i\phi_s}\pm p_ie^{i\phi_i}$, which gives $p_{s,i}=\sqrt{p_+^2+p_-^2\pm2p_+p_-e^{i(\phi_+-\phi_-)}}$, and \\
$h_2({p}_+,{p}_-,\phi_+-\phi_-,t)\equiv h_s(p_s,t)p_s^{-l_s}\ h_i(p_i,t)p_i^{-l_i}$.

\vspace{0.08in}
Eq. (\ref{tpda2pp}) is similar to Eq. (\ref{tpda2p}) where $e^{il_{s}\phi_{s}}$ and $e^{il_{i}\phi_{i}}$ are associated with the $z$-components of the OAM of each photon in the two one-photon fields. Following the same rule, one can associate the term $e^{il_+\phi_+}\equiv e^{i(m+n_s+n_i)\phi_+}$ in Eq. (\ref{tpda2pp}) with the $z$-component of the {\em intrinsic} part of OAM ($l_+\hbar$ for two photons) carried by two photons in the degrees of freedom of center-of-momentum movement (described by ${\bf p}_+$ and ${\bf q}_+$), and connect $e^{il_-\phi_-}\equiv e^{i[(l_s+l_i)-(m+n_s+n_i)]\phi_-}$ to the {\em extrinsic} part of OAM ($l_-\hbar$) of the two photons in the degrees of freedom of relative movement (described by ${\bf p}_-$ and ${\bf q}_-$) of one photon with respect to the other. Obviously, $l_++l_-=l_s+l_i$ always holds, which shows that the total OAM of the two photons can always be decomposed into two separate parts, the intrinsic part and the extrinsic part, in two independent sets of degrees of freedom of joint movement, proving the statements given in the introduction. In principle, neither the intrinsic OAM nor the extrinsic OAM is negligible when one considers the total OAM of two beams.

\vspace{0.0in}
\section{\label{sec:theory2} Extrinsic OAM of light in type-II SPDC processes}
For a type-I SPDC process, where the rotational symmetry around pump direction holds \cite{mair01}, the state vector of the down-converted light is calculated in \cite{arnaut00} to the first-order approximation. Here we consider the general case, in which the rotational symmetry may be broken in the SPDC processes. If one assumes a classical pump beam and two linearly-polarized down-converted modes ({signal} and {idler}) that are initially empty, the Hamiltonian governing a (type-I or type-II) SPDC process in the interaction picture is \cite{louisell61,hong85,arnaut00}
	\begin{eqnarray}
	\label{eq:ham}
\hat{H}_{\rm int} & = & \sum\limits_{{\bf k}_s,{\bf k}_i} {C}^{(s,i)}_j \int_{V_I}d^3r {E}_j({\bf r},t)e^{-i({\bf k}_s+{\bf k}_i)\cdot {\bf r}}\nonumber \\
 & & \times {\hat{a}}^\dagger({\bf k}_s){\hat{a}}^\dagger({\bf k}_i) e^{i(\omega_s+\omega_i)t} + {\rm H.C.}\, ,
	\end{eqnarray}
where $V_I$ is the nonlinear interaction volume, and ${\bf E}({\bf r},t)$ represents the electrical field associated with the pump beam. Subscripts {\em s} and {\em i} denote {signal} and {idler}, respectively. ${\hat{a}}^\dagger({\bf k}_s)$ and ${\hat{a}}^\dagger({\bf k}_i)$ are the creation operators for the down-converted modes and their wave vectors ${\bf k}_{s,i}$ are evaluated inside the medium. The coefficient ${C}^{(s,i)}_jV_I= -\sqrt{\hbar \omega_{s}/2 \varepsilon({\bf k}_s)}\ \sqrt{\hbar \omega_{i}/2 \varepsilon({\bf k}_i)}\ \chi_{jmn}({\bf e}_s)_m({\bf e}_i)_n$, where $\chi$ is the second-order nonlinearity tensor of the interaction medium and ${\bf e}$ represents the unit vector of linear polarization for the electrical field.

We consider a Laguerre-Gaussian (LG) pump beam propagating along $\hat{z}$ with the principal component polarized along $\hat{x}$ in cylindrical coordinates \cite{arnaut00,note2}
	\begin{eqnarray}
	\label{eq:pump}
{\bf E}(q_\rho,\phi,z;t) \equiv \psi^{(lp)}({\bf r})e^{i(k_{\mbox{\tiny {\it P}}}z-\omega_{\mbox{\tiny {\it P}}}t)}\hat{x}\, ,
	\end{eqnarray}
	\begin{eqnarray}
\psi^{(lp)}({\bf r})& = &\frac{A^{(lp)}}{\sqrt{1+(\frac{{\mbox{\small {\it z}}}}{{\mbox{\small {\it z}}}_{\mbox{\tiny {\it R}}}})^2}} \left[ \frac{\sqrt{2}q_\rho}{w(z)}\right]^lL^l_p\left[\frac{2q_\rho^2}{w^2(z)}\right]\nonumber\\ 
&  \times & e^{i{\mbox{\small {\it k}}}_{\mbox{\tiny {\it P}}}\frac{q_\rho^2}{2q(z)}}e^{il\tan^{-1}({\mbox{\small {\it y}}}/{\mbox{\small {\it x}}})} e^{-i(2p+l+1)\tan^{-1}({\mbox{\small {\it z}}}/{\mbox{\small {\it z}}}_{\mbox{\tiny {\it R}}})} \nonumber\, .
	\end{eqnarray}
Here ${\mbox{{\it z}}}_{\mbox{\tiny {\it R}}}$ is the Rayleigh length, $w(z)=w_0\sqrt{1+z^2/z^2_R}$, $w_0$ is the beam radius at the waist $z=0$. $l$ is the winding number of the pump mode and $p$ the number of radial nodes. Subscript $P$ refers to pump beam and $q(z)=z-i{\mbox{{\it z}}}_{\mbox{\tiny {\it R}}}$. Plugging Eq. (\ref{eq:pump}) into Eq. (\ref{eq:ham}) gives
	\begin{eqnarray}
	\label{eq:ham1}
\hat{H}_{\rm int} & = & \sum\limits_{{\bf p}_s,{\bf p}_i,\omega_s,\omega_i} {C}_1^{(s,i)} \tilde{\psi}^{(lp)}(\Delta {\bf k}) e^{i(\omega_s+\omega_i-\omega_{\mbox{\tiny {\it P}}})t} \nonumber \\
& & \times {\hat{a}}^\dagger({\bf p}_s,\omega_s){\hat{a}}^\dagger({\bf p}_i,\omega_i) + {\rm H.C.}\, ,
	\end{eqnarray}
where $\tilde{\psi}^{(lp)}(\Delta {\bf k})=\int_{V_I}d^3r{\psi}^{(lp)}({\bf r}){\mbox{exp}}(-i\Delta {\bf k}\cdot{\bf r})$, $\Delta {\bf k}={\bf k}_s+{\bf k}_i-{\bf k}_{\mbox{\tiny {\it P}}}$. Eq. (\ref{eq:ham1}) was relabeled ($\sum\limits_{\bf k}\rightarrow\sum\limits_{{\bf p},\omega}$) with beam invariants \cite{rubin96,pittman96} that are constant along propagation of beams: the angular frequencies $\omega_{s,i}$ and the transverse components ${\bf p}_{s,i}$ of wave vectors ${\bf k}_{s,i}$.

Under the assumption that the average radius of the beam is small compared to the transverse section of the nonlinear medium and that the medium length $l_c$ is much smaller than the Rayleigh range ${\mbox{{\it z}}}_{\mbox{\tiny {\it R}}}$ of the pump beam, we obtain \cite{arnaut00, barbosa02}
	\begin{eqnarray}
	\label{eq:geo}
\tilde{\psi}^{(lp)}(\Delta {\bf k}) = B^{(lp)}W(\Delta k_z)\varphi^{(lp)}(p_{+}) p^l_{+} e^{il\phi_{+}}\, ,
	\end{eqnarray}
where $B^{(lp)}=A^{(lp)}({\mbox{{\it z}}}_{\mbox{\tiny {\it R}}}\pi/k^{l+1}_{\mbox{\tiny {\it P}}})(\sqrt{2}{\mbox{{\it z}}}_{\mbox{\tiny {\it R}}}/w_0)^l{\mbox{exp}}[-\frac{\pi}{2}i(1-l-p)]2^{p-l+1}$, $W(\Delta k_z)=l_c\hspace{0.02in}\mbox{sinc}(\Delta k_zl_c/2)$, and 
	\begin{eqnarray}
	\label{eq:core}
\varphi^{(lp)}(p_{+}) = L_p^l\left(\frac{{\mbox{{\it z}}}_{\mbox{\tiny {\it R}}}}{k_{\mbox{\tiny {\it P}}}} p^2_{+}\right){\mbox{exp}}\left(-\frac{{\mbox{\small {\it z}}}_{\mbox{\tiny {\it R}}}}{2k_{\mbox{\tiny {\it P}}}}p^2_{+}\right)\, .
	\end{eqnarray}
We note that the spatial symmetry of the Hamiltonian (\ref{eq:ham1}) is primarily dictated by the term $W(\Delta k_z)$ through Eq. (\ref{eq:geo}).

Then, the two-photon wave function of the down-converted light reads \cite{arnaut00}, to the first-order approximation,
	\begin{eqnarray}
	\label{eq:twowave3}
|\psi(t)\rangle = |0\rangle &+& \sum\limits_{{\bf p}_s,{\bf p}_i,\omega_s,\omega_i} {C}_1^{(s,i)} T(\Delta \omega)\tilde{\psi}^{(lp)}(\Delta {\bf k})\nonumber \\
& & \times {\hat{a}}^\dagger({\bf p}_s,\omega_s){\hat{a}}^\dagger({\bf p}_i,\omega_i)|0\rangle\, ,
	\end{eqnarray}
where  $\Delta \omega =\omega_s+\omega_i-\omega_{\mbox{\tiny {\it P}}}$, $T(\Delta \omega)={\mbox{exp}}[i\Delta \omega(t-t_{\rm int}/2)]\sin(\Delta \omega t_{\rm int}/2)/(\Delta \omega/2)$ with $t_{\rm int}$ being the interaction time. Using equation (\ref{eq:twowave3}), we find the two-photon detection amplitude of the down-converted beams:
	\begin{eqnarray}
\varphi_2({\bf r}_s,{\bf r}_i)&\equiv&\langle0|\hat{E}_s^{(+)}({\bf r}_s)\hat{E}_i^{(+)}({\bf r}_i)|\psi(t)\rangle\nonumber\\
&=&\sum\limits_{{\bf p}_{s},{\bf p}_{i},\omega_s,\omega_i} C_{k_s}C_{k_i}{C}_1^{(s,i)} T(\Delta \omega)\tilde{\psi}^{(lp)}(\Delta {\bf k})\nonumber\\
&\times& e^{i({\bf k}_{s}\cdot{\bf r}_s+{\bf k}_{i}\cdot{\bf r}_i-\omega_st_s-\omega_it_i)}\nonumber\, ,
	\end{eqnarray}
 where $C_k=\sqrt{\frac{\hbar \omega_k}{2 \varepsilon_0 V}}$ and $V$ is the quantization volume. Using Eqs. (\ref{eq:geo}-\ref{eq:core}) and converting the sum into integrals $\sum\limits_{{\bf p},\omega}\equiv\sum\limits_{\bf k}\rightarrow$ $\frac{V}{(2\pi)^3}\int d^3k\rightarrow\frac{V}{(2\pi)^3}\int d^2p d\omega\frac{\omega}{c^2k_z}$ \cite{rubin96}, we obtain the transverse profile of $\varphi_2({\bf r}_s,{\bf r}_i)$ in the transverse planes $z_s=z_{0,s}$ and $z_i=z_{0,i}$:
	\begin{eqnarray} \label{tpda0}
\varphi_2({\bf q}_s,{\bf q}_i)&=&\frac{V^2}{(2\pi)^6}\int d^2p_{s}d^2p_{i}d\omega_sd\omega_i\frac{\omega_s}{c^2k_{z,s}}\frac{\omega_i}{c^2k_{z,i}}\nonumber\\
&\times& C_{k_s}C_{k_i}{C}_1^{(s,i)} T(\Delta \omega)B^{(lp)}W(\Delta k_z) p^l_{+} e^{il\phi_{+}}\nonumber\\
&\times& L_p^l\left(\frac{{\mbox{{\it z}}}_{\mbox{\tiny {\it R}}}}{k_{\mbox{\tiny {\it P}}}} p^2_{+}\right){\mbox{exp}}\left(-\frac{{\mbox{\small {\it z}}}_{\mbox{\tiny {\it R}}}}{2k_{\mbox{\tiny {\it P}}}}p^2_{+}\right)\nonumber\\
&\times& e^{i({\bf k}_{s}\cdot{\bf r}_s+{\bf k}_{i}\cdot{\bf r}_i-\omega_st_s-\omega_it_i)}\nonumber\\
&=&\int d^2p_{s}d^2p_{i}e^{i({\bf p}_{s}\cdot{\bf q}_s+{\bf p}_{i}\cdot{\bf q}_i)}\nonumber \\
&\times& B^{(lp)} L_p^l\left(\frac{{\mbox{{\it z}}}_{\mbox{\tiny {\it R}}}}{k_{\mbox{\tiny {\it P}}}} p^2_{+}\right)e^{-\frac{{\mbox{\small {\it z}}}_{\mbox{\tiny {\it R}}}}{{\mbox{\small 2{\it k}}}_{\mbox{\tiny {\it P}}}}p^2_{+}}p^l_{+}e^{il\phi_+}\nonumber \\
&\times& \frac{V^2}{(2\pi)^6} \int d\omega_s d\omega_i \frac{C_{k_s}C_{k_i}\omega_s\omega_i}{c^4k_{z,s}k_{z,i}} {C}_1^{(s,i)} T(\Delta \omega)\nonumber \\
&\times& W(\Delta k_z) e^{i(k_{z,s}z_{0,s}-\omega_st_s+k_{z,i}z_{0,i}-\omega_it_i)} \, .
	\end{eqnarray}

For the sake of simplicity, we assume $p_s\approx p_i$ \cite{rubin96} and $z_{0,s}=z_{0,i}\equiv z'_0$, $t_s=t_i\equiv t$, $\bar{\omega}_s=\bar{\omega}_i\equiv\bar{\omega}$ (frequency-degenerate case), where $\bar{\omega}_{s,i}$ are the central angular frequencies of the down-converted beams. Usually, $T(\Delta \omega)$ can be approximated by a Delta function $\delta(\Delta \omega)$ times $t_{\rm int}$ and, under paraxial approximation, the phase factor $\phi_{s,i}\equiv k_{z,s}z_{0,s}-\omega_st_s+k_{z,i}z_{0,i}-\omega_it_i$ may be considered as a constant over the integral range of the angular frequency:
	\begin{eqnarray}
\delta\phi_{s,i}&\approx&(\frac{z'_0}{c}-t)(\delta\omega_s+\delta\omega_i)+\frac{cz'_0}{2}(\frac{p_s^2}{\bar{\omega}^2}\delta\omega_s+\frac{p_i^2}{\bar{\omega}^2}\delta\omega_i)\nonumber\\
&\approx& \frac{cz'_0}{2}(\frac{p_s^2}{\bar{\omega}^2}-\frac{p_i^2}{\bar{\omega}^2})\delta\omega_s\approx 0\nonumber\, ,
	\end{eqnarray}
where $\delta(\omega_s+\omega_i)=0$ is applied. Then, with joint variables ${\bf p}_\pm$ and ${\bf q}_\pm$, Eq. (\ref{tpda0}) can be re-written as 
	\begin{eqnarray} \label{tpda1}
& & \varphi_2({\bf q}_+,{\bf q}_-)=\frac{V^2t_{\rm int}}{4(2\pi)^6}\int d^2p_{+}e^{i(\frac{{\bf q}_+}{2}\cdot{\bf p}_{+}-\frac{cz'_0}{4\bar{\omega}}p^2_+)} F_+({\bf p}_+)\nonumber \\
&&\times \int d^2p_{-}e^{i(\frac{{\bf q}_-}{2}\cdot{\bf p}_{-}-\frac{cz'_0}{4\bar{\omega}}p^2_-)}\int d\omega_s D(\omega_s) W(\Delta k_z)\, ,
	\end{eqnarray}
where
	\begin{eqnarray} \label{Fp}
&&F_+({\bf p}_+)= \left[B^{(lp)} p^l_{+}L_p^l\left(\frac{{\mbox{{\it z}}}_{\mbox{\tiny {\it R}}}}{k_{\mbox{\tiny {\it P}}}} p^2_{+}\right)e^{-\frac{{\mbox{\small {\it z}}}_{\mbox{\tiny {\it R}}}}{{\mbox{\small 2{\it k}}}_{\mbox{\tiny {\it P}}}}p^2_{+}}\right]e^{il\phi_+}, \\
&&D(\omega_s)=\frac{C_{k_s}C_{k_i}\omega_s\omega_i}{c^4k_{z,s}k_{z,i}} {C}_1^{(s,i)}|_{\omega_i=\omega_{\mbox{\tiny {\it P}}}-\omega_s}\nonumber \, ,
	\end{eqnarray}
with the global phase term $e^{i\omega_{\mbox{\tiny {\it P}}}(\frac{z'_0}{c}-t)}$ being dropped off. The dependence of $D(\omega_s)$ on ${\bf p}_\pm$ is considered weak and neglected in our analysis.

In terms of the beam invariants, we evaluate the phase mismatch $\Delta k_z$ to the first-order approximation: \cite{rubin96}
	\begin{eqnarray}\label{deltak}
\Delta k_z &\approx& -\bar{\nu} D - \frac{p_{+}^2+p_{-}^2}{4\bar{K}} - \frac{N}{2} ({\bf p}_{+}-{\bf p}_{-})\cdot \hat{x} \nonumber \\
 &\approx& -\bar{\nu} D - \frac{p_{-}^2}{4\bar{K}} + \frac{N}{2} {\bf p}_{-}\cdot \hat{x} \, ,
	\end{eqnarray}
where the {signal} is assumed to be the {\em e}-beam. Here 
$\bar{\nu}$, $D$, $\bar{K}$, and $N$ are parameters dependent of the nonlinear medium and defined in Ref. \cite{rubin96}. In the last step, it is assumed that $|{\bf p}_{+}| \ll |{\bf p}_{-}|$, which is usually valid in non-collinear configurations except in a very rare case where the detected down-converted photon-pairs nearly co-propagate with the pump. In this case, the dependence of $W(\Delta k_z)$ on ${\bf p}_{+}$ in Eq. (\ref{tpda1}) is negligible. At this point, one might argue that an important and largely employed experimental configuration is the non-collinear phase matching in which the x-components of ${\bf p}_{+}$ and ${\bf p}_{-}$ are comparable, that is, where the two down-converted light cones cross each other. Now we show that the ${\bf p}_+ \cdot \hat{x}$ term can be neglected even if ${\bf p}_+ \cdot \hat{x}$ is comparable to ${\bf p}_- \cdot \hat{x}$.

At the crossings of the two down-conversion cones, both ${\bf p}_\pm \cdot \hat{x}$ may be comparable to each other because they are close to zero with respect to other cases where the two cones do not cross. So, both terms are small compared to the $|{\bf p}_-|$ term and can be omitted in Eq. (\ref{deltak}). To quantitatively prove this, we take the experimental example of Ref. \cite{kwiat95} , where the effective $\theta_{\mbox{pm}}=49.63^\circ$. One can easily obtain $\bar{K}=14.38\mu \mbox{m}^{-1}$, $N=-0.068$ using the formula in \cite{rubin96}. According to the numerical estimation in \cite{rubin96}, $|{\bf p}_-|$ is order of $1\mu \mbox{m}^{-1}$ at the crossings. So,
\begin{eqnarray}
\frac{p_-^2}{4\bar{K}}\approx \frac{0.035}{2}p_-, \ \mbox{and}\ \frac{N}{2}{\bf p}_+ \cdot \hat{x}\approx 0.034{\bf p}_+ \cdot \hat{x}\nonumber\, .
\end{eqnarray}
Then we can do the following comparison for the cone crossings:
\begin{eqnarray}
|{\bf p}_+ \cdot \hat{x}| \le p_+ << p_- \  \rightarrow \ \frac{N}{2}|{\bf p}_+ \cdot \hat{x}|<<\frac{p_-^2}{4\bar{K}}\nonumber\, .
\end{eqnarray}
Because $|{\bf p}_+ \cdot \hat{x}|\approx |{\bf p}_- \cdot \hat{x}|$ at the crossings, we also have 
\begin{eqnarray}
\frac{N}{2}|{\bf p}_- \cdot \hat{x}|<<\frac{p_-^2}{4\bar{K}}\nonumber\, .
\end{eqnarray}

Mathematically, it does not hurt to keep one negligible term and drop the other negligible one in Eq. (\ref{deltak}). Choosing to keep the ${\bf p}_- \cdot \hat{x}$ term is to make Eq. (\ref{deltak}) general enough to cover all cases no matter whether ${\bf p}_\pm \cdot \hat{x}$ are comparable to each other or not.

Accordingly, the two-photon detection amplitude $\varphi_2({\bf q}_+,{\bf q}_-)$ can be written as a product of two separate terms:
	\begin{eqnarray}\label{tpda4}
\varphi_2({\bf q}_+,{\bf q}_-)&=& R_+({\bf q}_+)\hspace{0.05in}R_-({\bf q}_-)\, ,
	\end{eqnarray} 
where
	\begin{eqnarray} \label{Rp}
&&R_+({\bf q}_+)= \int d^2p_{+}e^{i({\bf p}_{+}\cdot\frac{{\bf q}_+}{2}-p^2_+\frac{cz'_0}{4\bar{\omega}})} F_+({\bf p}_+),\\\label{Rm}
&&R_-({\bf q}_-)= \int d^2p_{-}e^{i({\bf p}_{-}\cdot\frac{{\bf q}_-}{2}-p^2_-\frac{cz'_0}{4\bar{\omega}})} F_-({\bf p}_-),\\\label{Fm}
&&F_-({\bf p}_-)= \frac{V^2t_{\rm int}}{4(2\pi)^6} \int d\omega_s D(\omega_s) W[\Delta k_z(\omega_s,{\bf p}_-)]\, .
	\end{eqnarray}
Eq. (\ref{tpda4}) can be generalized to the type-I case and its Fourier transform reads \cite{note1}: 
	\begin{eqnarray}\label{tpda5}
F_2({\bf p}_+,{\bf p}_-)&=& F_+({\bf p}_+)\hspace{0.05in}F_-({\bf p}_-)\, ,
	\end{eqnarray}
if one denotes $F_2({\bf p}_+,{\bf p}_-)$ as the Fourier transform of $\varphi_2({\bf q}_+,{\bf q}_-)$.

Eq. (\ref{tpda4}) or Eq. (\ref{tpda5}) reveals the following physics. The movement of the down-converted photon-pairs in the transverse plane is de-coupled into {\em two} independent sets of degrees of freedom: one for the {\em center-of-momentum} movement described with the joint variables $\{{\bf q}_+, {\bf p}_+\}$ and the other for the {\em relative} movement of each photon with respect to its twin delineated with $\{{\bf q}_-, {\bf p}_-\}$. In the degrees of center-of-momentum-movement freedom, the photon-pairs carry intrinsic OAM (denoted as $l_+\hbar$ per pair) always equal to that of the pump photon, i.e., $l_+=l$, as stated by Eq. (\ref{Fp}), giving rise to entanglement in intrinsic OAM in these degrees of freedom. In the degrees of relative-movement freedom, as attested by Eq. (\ref{Rm})-(\ref{Fm}) and shown below in details, the extrinsic OAM ($l_-\hbar$ per pair) carried by the photon-pairs depends on the azimuthal symmetry of $F_-({\bf p}_-)$ dictated by $W[\Delta k_z(\omega_s,{\bf p}_-)]$.

   \begin{figure}[htb]
   \begin{center}
\centerline{\hspace{0.2in}\scalebox{0.175}{\includegraphics{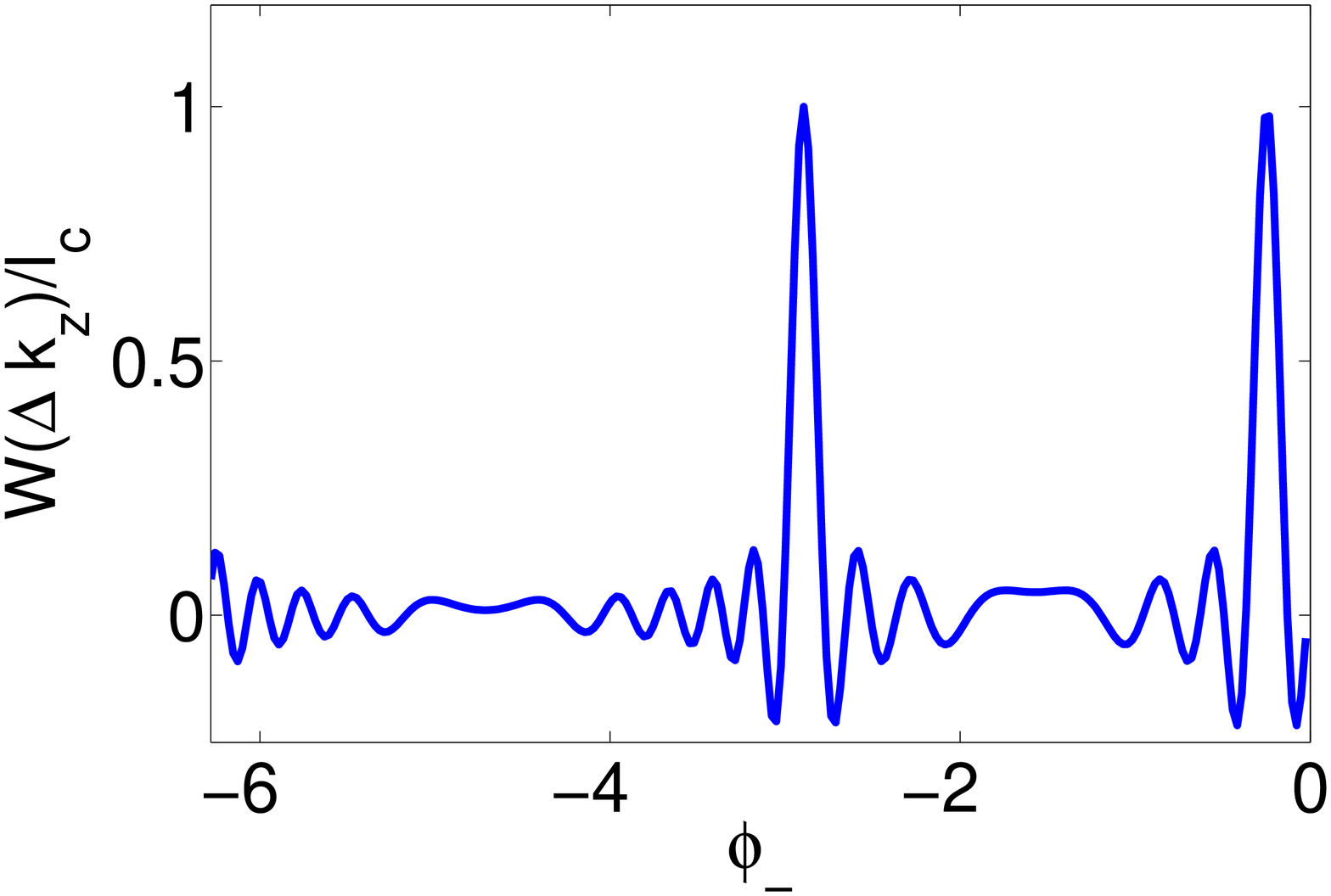}}
            \hspace{-0.15in}\scalebox{0.175}{\includegraphics{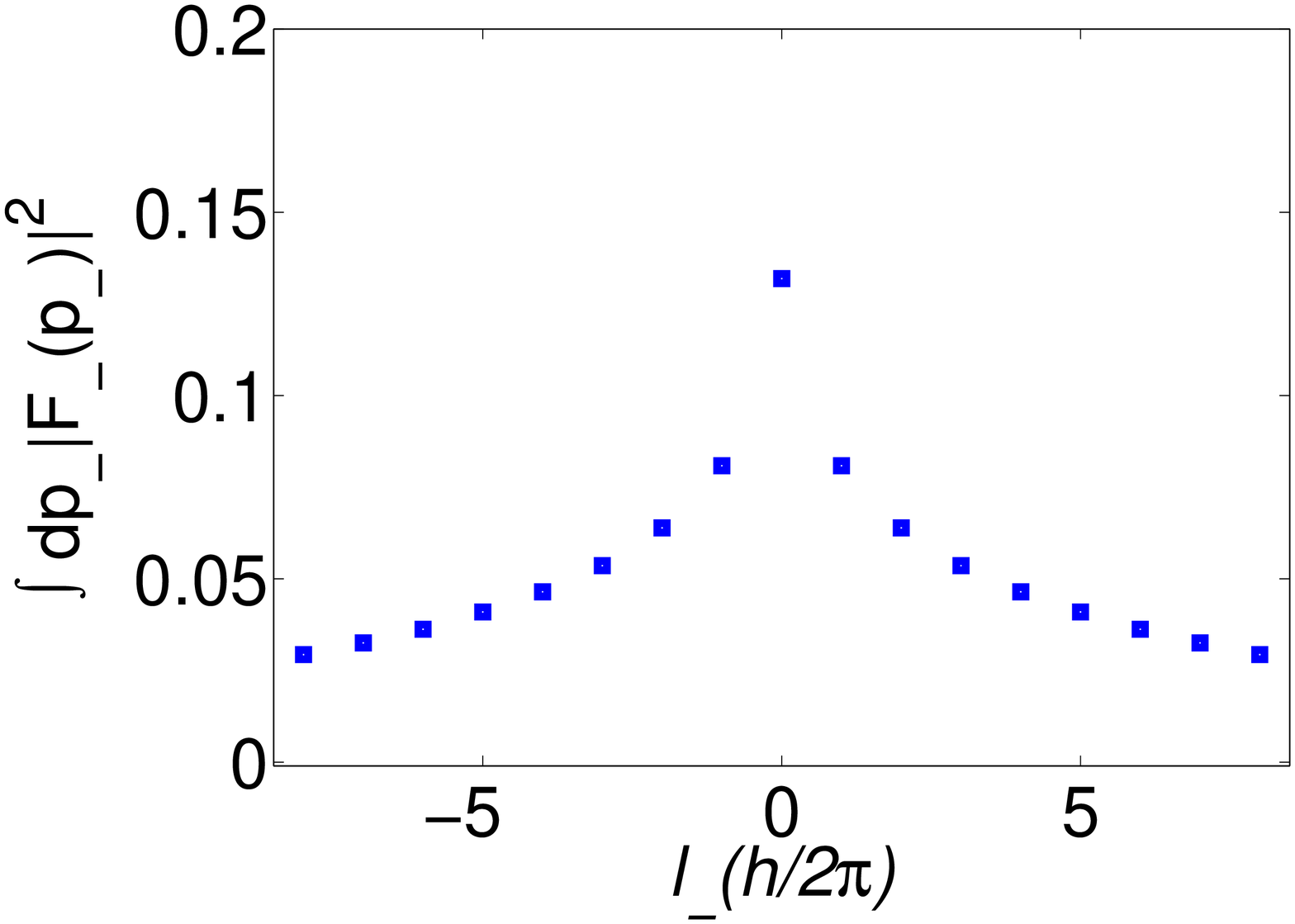}}\hspace{0.05in}}
\vspace{-1.175in}
\centerline{\hspace{0.2in}(a)\hspace{1.45in}(b)\hspace{0.7in}}
\vspace{0.615in}
   \end{center}
   \caption[Phase mismatch of type-II SPDC ] 
   {\label{pic:ccim} (color online) The spatial asymmetry in a type-II SPDC process, where $l_c=$ 0.5mm, and the non-negligible extrinsic OAM of the down-converted beams. (a) Typical azimuthal-angle dependence of $W(\Delta k_z)l_c^{-1}$, which is a measure for the degree of spatial symmetry breaking of the Hamiltonian. (b) The (normalized) probability of generating down-converted photon-pairs as a function of extrinsic OAM $l_-\hbar$ per pair in the degrees of relative-movement freedom.
}
   \end{figure} 

Mathematically, one can always expand Eq. (\ref{Fm}) in the form of Fourier series:
	\begin{eqnarray} \label{Fmexp}
&&F_-({\bf p}_-)= \sum\limits_m F^{(m)}_-(p_-)e^{im\phi_{-}}\, ,
	\end{eqnarray}
where each non-zero $F^{(m)}_-(p_-)$ term gives rise to a probability $\propto \int dp_-|F^{(m)}_-(p_-)|^2$ that the down-converted photon-pairs carry extrinsic OAM of $m\hbar$ ($l_-=m$) per pair in the degrees of relative-movement freedom, according to the preceding section [Eq. (\ref{tpda2pp})]. In a type-II SPDC process, $W[\Delta k_z(\omega_s,{\bf p}_-)]$ (and then the Hamiltonian $\hat{H}_{\rm int}$) lacks the azimuthal symmetry in usual experimental conditions $l_c\ge 0.5$mm \cite{rubin96} [Fig. \ref{pic:ccim}(a)]. As stated by Eq. (\ref{Fm}), $F_-({\bf p}_-)$ then must be a function of the azimuthal angle $\phi_-$, which requires that at least one higher-order term in Eq. (\ref{Fmexp}) is non-zero. Each of these non-zero higher-order terms ($l_-=m\ne 0$) contributes non-negligible extrinsic OAM of the down-converted photon-pairs in the degrees of relative-movement freedom [Fig. \ref{pic:ccim}(b)]. 

On the contrary, the phase mismatch in type-I SPDC processes is, to the first-order approximation, \cite{rubin96}
	\begin{eqnarray}
\Delta k_z &\approx& -\bar{\nu} D - \frac{p_{+}^2+p_{-}^2}{4\bar{K}} \nonumber \\
 &\approx& -\bar{\nu} D - \frac{p_{-}^2}{4\bar{K}} \nonumber\, ,
	\end{eqnarray}
where non-collinear configurations are assumed. Obviously, the phase mismatch in the type-I cases is azimuthally invariant. Thereby, in the Fourier expansion of Eq. (\ref{Fmexp}), only the zero-order term survives with all the higher-order terms being left null. In other words, the down-converted photon-pairs are generated in type-I SPDC processes with 100\% probability that they carry no extrinsic OAM, in principle, due to the azimuthal symmetry of the non-linear processes.

\section{\label{sec:discussion} Discussions}

In the foregoing section, we theoretically address that the down-converted beams created in type-II SPDC processes carry non-negligible extrinsic OAM in the degrees of freedom of relative movement. Due to the misunderstanding of the term of thin-medium approximation, the existence of the extrinsic portion of the OAM in the type-II SPDC processes has been theoretically overlooked for years \cite{arnold02,walborn04}.

\begin{figure}[htb]
   \begin{center}
\centerline{\ }
\centerline{\hspace{-0.1in}\scalebox{0.14}{\includegraphics{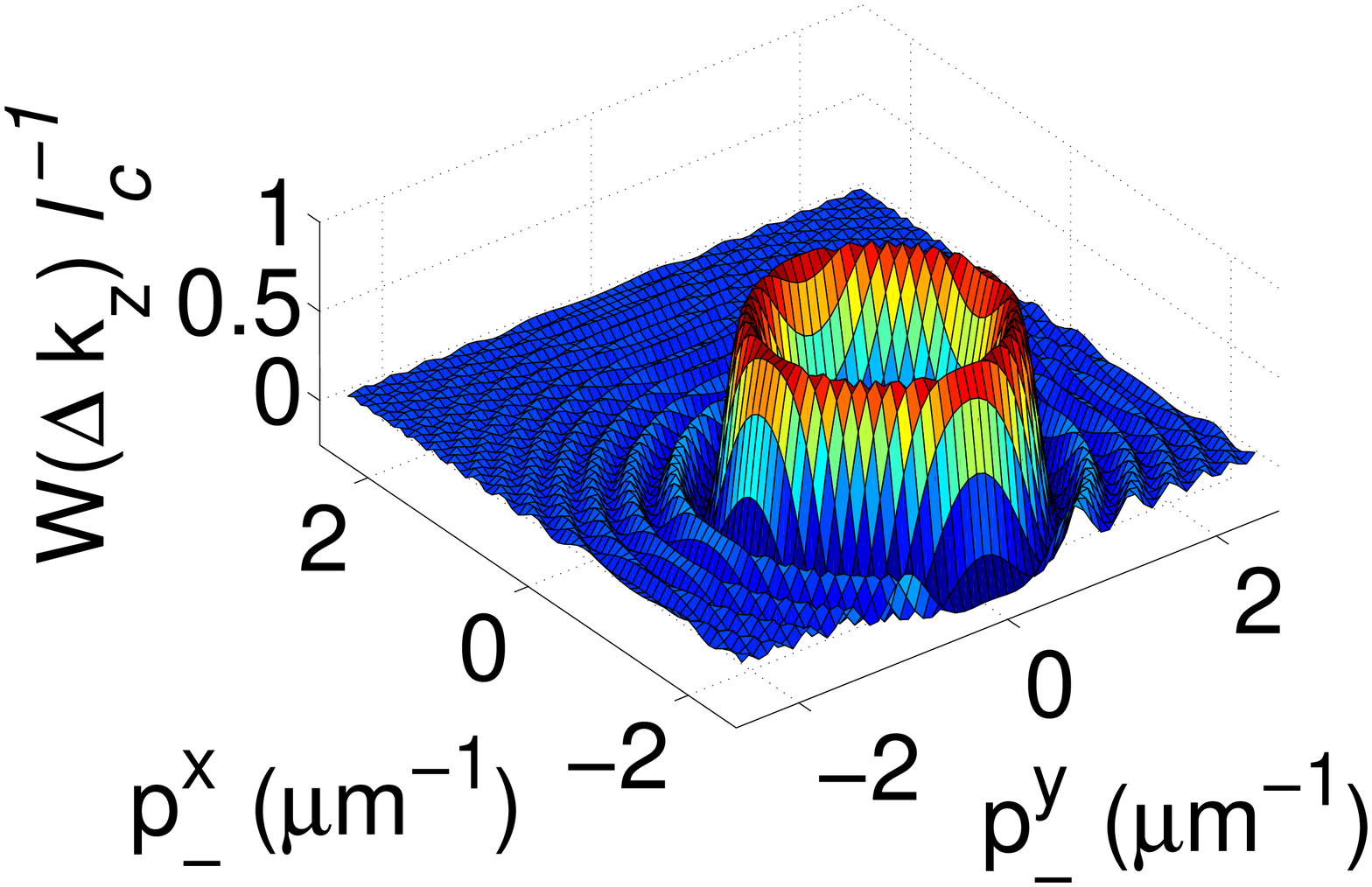}}\hspace{0.3in}
            \scalebox{0.14}{\includegraphics{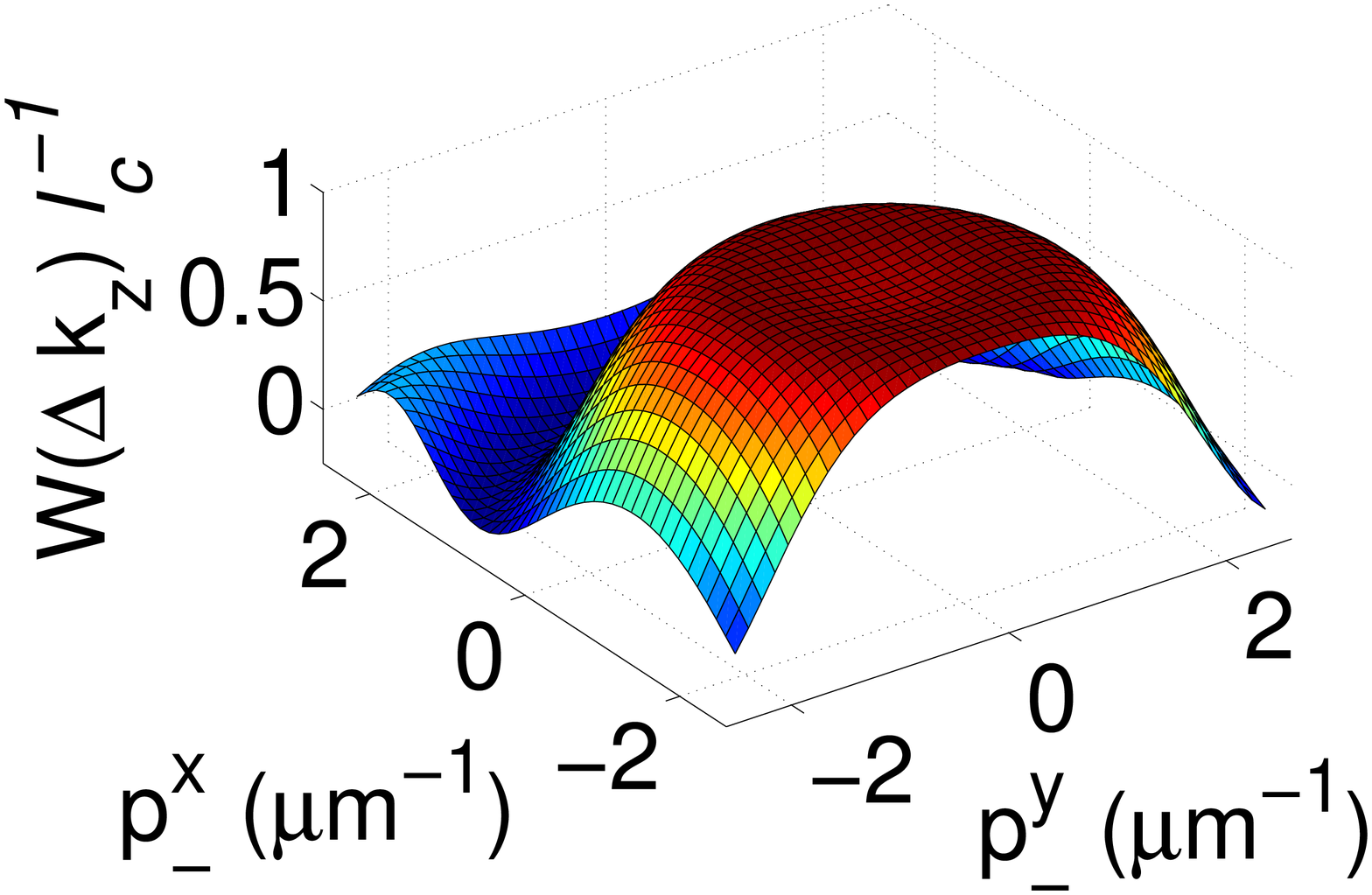}}\hspace{0.1in}}
\vspace{-0.97in}
\centerline{\hspace{0.6in}(a)\hspace{1.60in}(b)\hspace{1.4in}}
\vspace{0.5in}
   \end{center}
   \caption[Phase mismatch of type-II SPDC ] 
   {\label{pic:pmm} (color online) $W(\Delta k_z)l_c^{-1}$ for type-II SPDC processes. (a) $l_c=0.1$mm. (b) $l_c=10\mu$m.
}
   \end{figure}

In previous theoretical studies \cite{arnold02,walborn04}, $W(\Delta k_z)l_c^{-1}$, denoted as $\Delta({\bf p}_-)$ in \cite{arnold02}, was either approximated by one \cite{walborn04} or considered as {\em a very broad function that cuts out modes with large transverse wave vectors} \cite{arnold02}. The treatments for $W(\Delta k_z)l_c^{-1}$ in these theories are valid only in thin-medium approximation, which demands medium thickness to be order of $10\mu$m, as shown in Fig. \ref{pic:pmm}. In this case, $W(\Delta k_z)l_c^{-1}$ is approximately azimuthally symmetric even for type-II SPDC processes. However, in practice, all experiments involving SPDC processes are far beyond valid thin-medium approximation and the $W(\Delta k_z)l_c^{-1}$ does not necessarily possess azimuthal symmetry [Fig. \ref{pic:ccim}(a)].

Concerning experimental investigations, one might wonder why the extrinsic OAM of the down-converted beams in the type-II SPDC processes has never been observed. The very reason is that the existing OAM measurement techniques \cite{mair01,Oemrawsingh04ao} are not suitable for measuring the extrinsic part of the OAM of the down-converted beams. To see this, let us first exam how the traditional scheme works for intrinsic OAM measurement.

\begin{figure}[htb]
   \begin{center}
\centerline{\scalebox{0.25}{\includegraphics{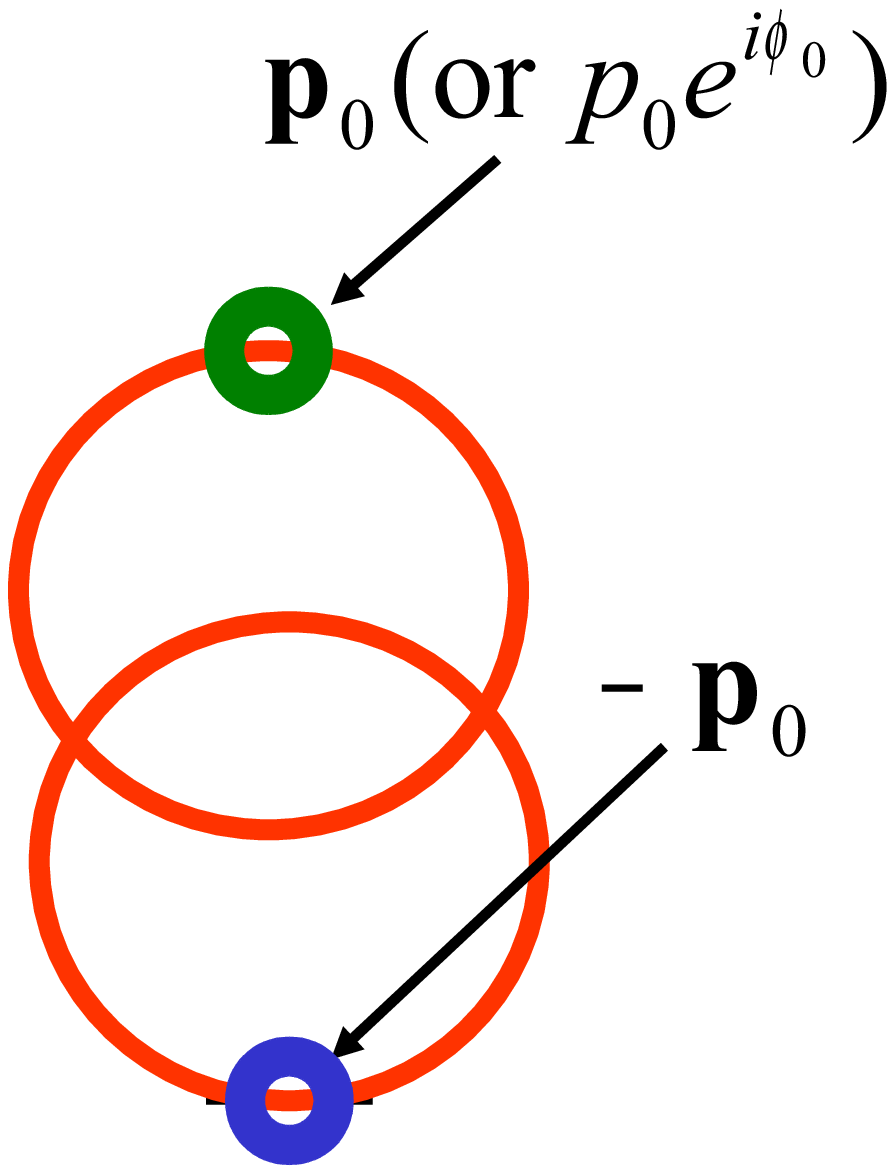}}\hspace{0.5in}
            \scalebox{0.17}{\includegraphics{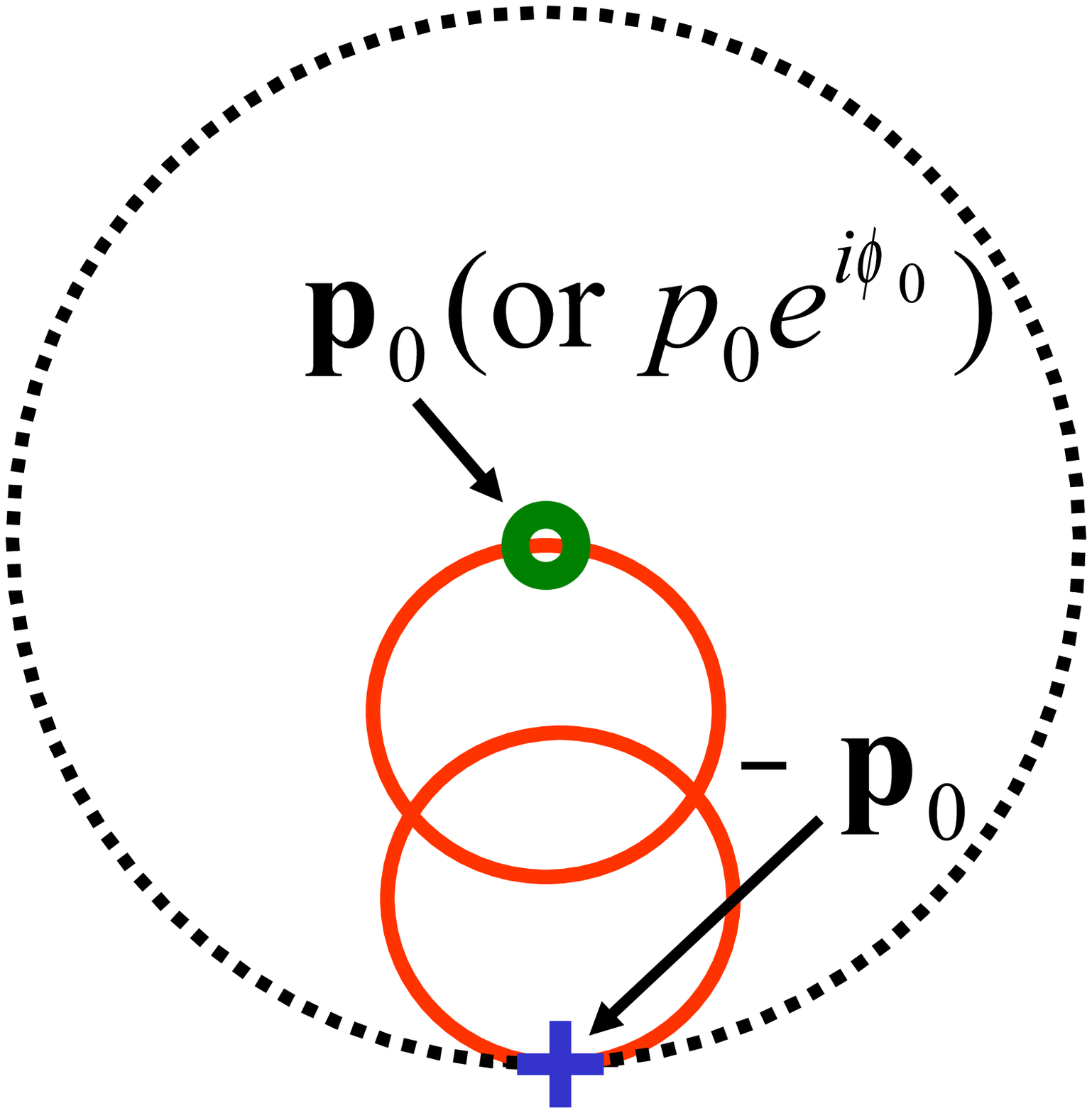}}\hspace{0.0in}}
\vspace{0.1in}
\centerline{\hspace{1.2in}(a)\hspace{1.55in}(b)\hspace{1.4in}}
\vspace{-0.3in}
   \end{center}
   \caption[Phase mismatch of type-II SPDC ] 
   {\label{pic:OAMmeas} (color online) (a)  Intrinsic OAM measurement in the traditional scheme \cite{mair01}, where two sets of optical devices are located at conjugate positions ($\pm{\bf p}_0$) on the down-conversion rings (red circles) to detect the donut-shaped mode of each down-converted beam. (b) Illustrating how the extrinsic OAM carried by photon-pairs escapes detection in the traditional OAM measurement scheme. In the degrees of relative-movement freedom, detection of one photon by one set of detection devices centered at ${\bf p}_0$ (donut-like marker) projects its twin into a mode (indicated by the dashed circle) centered also at ${\bf p}_0$, which is out of detection scope for the other set of devices centered at $-{\bf p}_0$ (blue cross).
}
   \end{figure}

One can expand Eq. (\ref{Fp}), which describes the center-of-momentum movement of the down-converted photon-pairs, as (each photon-pair carries intrinsic OAM $l\hbar$, i.e., $l_+=l$)
	\begin{eqnarray} \label{Fpexp1}
&&F_+({\bf p}_+)\equiv F_+^{(l)}(p_+)e^{il\phi_+}=\sum\limits_n F_+^{(l,n)} p_+^{n-l}(p_+e^{i\phi_+})^{l}\nonumber \\
&=&\left[\sum\limits_n F_+^{(l,n)} p_+^{n-l}\right]\left(p''_se^{i\phi''_s}+p''_ie^{i\phi''_i}\right)^{l}\nonumber \\
&=&\left[\ \cdot\ \right] \sum\limits_m \bin{l}{m} ({p''_s})^me^{im\phi''_s}({p''_i})^{l-m}e^{i(l-m)\phi''_i}\nonumber\, ,
	\end{eqnarray}
where ${p''}_{s} e^{i\phi''_{s}}={p}_{s}e^{i\phi_{s}}-p_0e^{i\phi_0}$ and ${p''}_{i} e^{i\phi''_{i}}={p}_{i}e^{i\phi_{i}}+p_0e^{i\phi_0}$ represent new vectors centered at $\pm{\bf p}_0$ (or $\pm p_0e^{i\phi_0}$, $\phi_0$ is the azimuthal angle of ${\bf p}_0$) respectively in the transverse planes. So, in the degrees of center-of-momentum-movement freedom, if the {signal} beam is projected into a mode centered at ${\bf p}_0$ carrying OAM $l_s\hbar$ ($m=l_s$) per photon, its twin beam will be simultaneously projected into another mode with a center of $-{\bf p}_0$ carrying OAM $(l-l_s)\hbar$ per photon, which can be measured by a phase mask centered at $-{\bf p}_0$ combined with an SMF \cite{mair01,Oemrawsingh05}, as is illustrated in Fig. \ref{pic:OAMmeas}a.

Similarly, in the degrees of relative-movement freedom, one can expand Eq. (\ref{Fmexp}) as (assuming each photon-pair carries extrinsic OAM $l'\hbar$, i.e., $l_-=l'$, in these degrees of freedom due to symmetry breaking)
	\begin{eqnarray} \label{Fmexp1}
&&F_-({\bf p}_-)\equiv F_-^{(l')}(p_-)e^{il'\phi_-}\nonumber \\
&=&\sum\limits_n F_-^{(l',n)} p_-^{n-l'}(p_-e^{i\phi_-})^{l'}\nonumber \\
&=&\left[\sum\limits_n F_-^{(l',n)} p_-^{n-l'}\right]\left(p'_se^{i\phi'_s}-p'_ie^{i\phi'_i}\right)^{l'}\nonumber \\
&=&\left[\ \cdot\ \right] \sum\limits_m \bin{l'}{m} (-1)^{l'-m}({p'_s})^me^{im\phi'_s}({p'_i})^{l'-m}e^{i(l'-m)\phi'_i}\nonumber\, ,
	\end{eqnarray}
where ${p'}_{s,i} e^{i\phi'_{s,i}}={p}_{s,i}e^{i\phi_{s,i}}-p_0e^{i\phi_0}$ represent new vectors both centered at ${\bf p}_0$ in the transverse planes. In the degrees of relative-movement freedom, if the {signal} beam is projected into a mode centered at ${\bf p}_0$ carrying OAM $l_s\hbar$ ($m=l_s$) per photon, its twin beam will be simultaneously projected into another mode with a center of ${\bf p}_0$ also, carrying OAM $(l'-l_s)\hbar$ per photon which nevertheless cannot be measured by a phase mask centered at $-{\bf p}_0$ combined with an SMF (Fig. \ref{pic:OAMmeas}b) that measures only the intrinsic part of OAM in the degrees of the center-of-momentum-movement freedom in the traditional scheme \cite{mair01,Oemrawsingh05}.

In conclusion, we theoretically show the existence of non-negligible extrinsic OAM carried by the down-converted beams generated in the type-II SPDC processes in the degrees of freedom of relative movement due to the azimuthal symmetry-breaking. We explain how the extrinsic OAM escapes detection in the traditional OAM measurement scheme. Therefore, new OAM measurement techniques need to be developed if the extrinsic OAM is to be experimentally studied in the SPDC processes.

\begin{acknowledgments}
This work was supported in part by the Quantum Imaging MURI funded through the U.S. Army Research Office. 
\end{acknowledgments}


\begin{thebibliography}{}
\bibitem{note2} 
The beam deformation due to the anisotropy of the medium is small and negligible in all practical SPDC experiments, as is confirmed by all published experimental results.
\end{thebibliography}
\end{document}